\documentclass[12pt]{iopart}
\pdfoutput=1

\usepackage{graphicx}
\usepackage{color}
\usepackage{iopams}
\usepackage{braket}
\usepackage{amssymb}
\usepackage{hyperref}

\begin{document}

\title{A single-atom quantum memory in silicon}

\author{Solomon Freer$^1$, Stephanie Simmons$^1$\footnote{Present address: Department of Physics, Simon Fraser University, Burnaby, British Columbia, Canada V5A 1S6.}, Arne Laucht$^1$, Juha T Muhonen$^1$\footnote{Present address: Center for Nanophotonics, FOM Institute AMOLF, Science Park 104, 1098 XG Amsterdam, The Netherlands.}, Juan P Dehollain$^1$\footnote{Present address: QuTech \& Kavli Institute of Nanoscience, TU Delft, 2628 CJ Delft, The Netherlands.}, Rachpon Kalra$^1$\footnote{Present address: School of Mathematics \& Physics, University of Queensland, Brisbane QLD 4072, Australia.}, Fahd A Mohiyaddin$^1$\footnote{Present address:  Quantum Computing Institute, Oak Ridge National Laboratory, Oak Ridge, Tennessee, USA.}, Fay E Hudson$^1$, Kohei M Itoh$^2$, Jeffrey C McCallum$^3$, David N Jamieson$^3$, Andrew S Dzurak$^1$ and Andrea Morello$^1$}
\address{$^1$ Centre for Quantum Computation and Communication Technology, School of Electrical Engineering \& Telecommunications, UNSW Australia, Sydney NSW 2052, Australia}
\address{$^2$ School of Fundamental Science and Technology, Keio University, 3-14-1 Hiyoshi, 223-8522, Japan}
\address{$^3$ Centre for Quantum Computation and Communication Technology, School of Physics, University of Melbourne, Melbourne VIC 3010, Australia}

\date{\today}% It is always \today, today,

%\date{}
%\renewcommand{\abstractname}{\vspace{-\baselineskip}}
\begin{abstract}
	Long coherence times and fast gate operations are desirable but often conflicting requirements for physical qubits. This conflict can be resolved by resorting to fast qubits for operations, and by storing their state in a `quantum memory' while idle. The $^{31}$P donor in silicon comes naturally equipped with a fast qubit (the electron spin) and a long-lived qubit (the $^{31}$P nuclear spin), coexisting in a bound state at cryogenic temperatures. Here, we demonstrate storage and retrieval of quantum information from a single donor electron spin to its host phosphorus nucleus in isotopically-enriched $^{28}$Si. The fidelity of the memory process is characterised via both state and process tomography. We report an overall process fidelity $F_p \approx 81$\%, a memory fidelity $F_m \approx 92$\%, and memory storage times up to 80 ms. These values are limited by a transient shift of the electron spin resonance frequency following high-power radiofrequency pulses.
\end{abstract}

\maketitle

%TODO: Z-tomo table
%In thesis, explain the amp/vis stuff, but not here
\section{Introduction}

In classical computation, memory is a trivial resource, implemented simply by making copies of digital information in long-lived memory hardware, and retrieving the information at later stages for further processing. In quantum computation, this is forbidden by the fundamental inability of arbitrary quantum states to be faithfully copied, as dictated by the no-cloning theorem \cite{Wootters1982}. Therefore, in the absence of additional resources or quantum error correction, the storage of a quantum bit (qubit) is limited by the intrinsic coherence time of the physical qubit that holds the information. `Quantum memories' \cite{Simon2010} surpass the storage limits of a given physical qubit by transferring the quantum information onto a `memory qubit' which is less susceptible to environmental noise and dephasing, then returning the information to the preferred `processing qubit' when required, once again allowing fast manipulation and/or readout. Alternatively, memory qubits can serve the purpose of storing a `flying qubit' such as a photon in a long-lived solid-state system \cite{Julsgaard2004,Wilk2007,Lvovsky2009,Hedges2010}.

The phosphorus donor in silicon provides a natural pair of qubits for these purposes. The $^{31}$P electron and nuclear spins are coupled to external magnetic fields via their gyromagnetic ratios $|\gamma_e| \approx 28.0$~GHz/T and $|\gamma_n| \approx 17.2$~MHz/T, respectively. The strong magnetic coupling of the electron enables fast quantum gate operations via magnetic resonance pulses \cite{Pla2012}, while the weaker coupling of the nucleus makes it highly insensitive to environmental noise, resulting in long coherence times \cite{Steger2012,Saeedi2013,Pla2013,Muhonen2014}. Donors in silicon have the further benefits of low spin-orbit coupling and dominance of spin-zero isotopes in the crystal lattice \cite{Witzel2010}, limiting electric and magnetic decoherence sources, respectively. In the present device, isotopic enrichment of spin-zero $^{28}$Si in the silicon substrate was used to further reduce magnetic decoherence \cite{Muhonen2014,Itoh2014}.

Recent experiments have shown the great potential of donor qubits for quantum information processing \cite{Pla2012,Dupont2013,Pla2013,Dehollain2014,Muhonen2014,Gonzalez2014,Weber2014,Dehollain2015,Laucht2015,Harvey2015,Urdampilleta2015}, and some new proposals for large-scale architectures based on $^{31}$P suggest the swapping of quantum states between electron and nucleus to optimize gate speeds and operation fidelities \cite{Tosi2015arxiv,Hill2015}. Storage of the classical \cite{McCamey2010} or the quantum \cite{Morton2008} state of P electron spins onto their $^{31}$P nuclei has been demonstrated in ensemble experiments. However, quantum memory demonstrations at the single-qubit level have so far been limited to trapped ions \cite{Kielpinski2001,Specht2011} and to defects in diamond \cite{Dutt2007,Fuchs2011,Maurer2012,Shim2013,Reiserer2016}.

Here we present the experimental demonstration of storage and retrieval of quantum information between the electron and the nucleus of a single $^{31}$P donor in isotopically enriched $^{28}$Si. For brief storage times we achieve an overall process fidelity $F_p = 81\pm 7$\%, well beyond the classical limit of $2/3$ \cite{Massar1995}, and we show that storage times can be extended up to 80 ms via dynamical decoupling (DD) of the nuclear spin memory.

\begin{figure}
\centering
\includegraphics[width=\columnwidth, keepaspectratio = true]{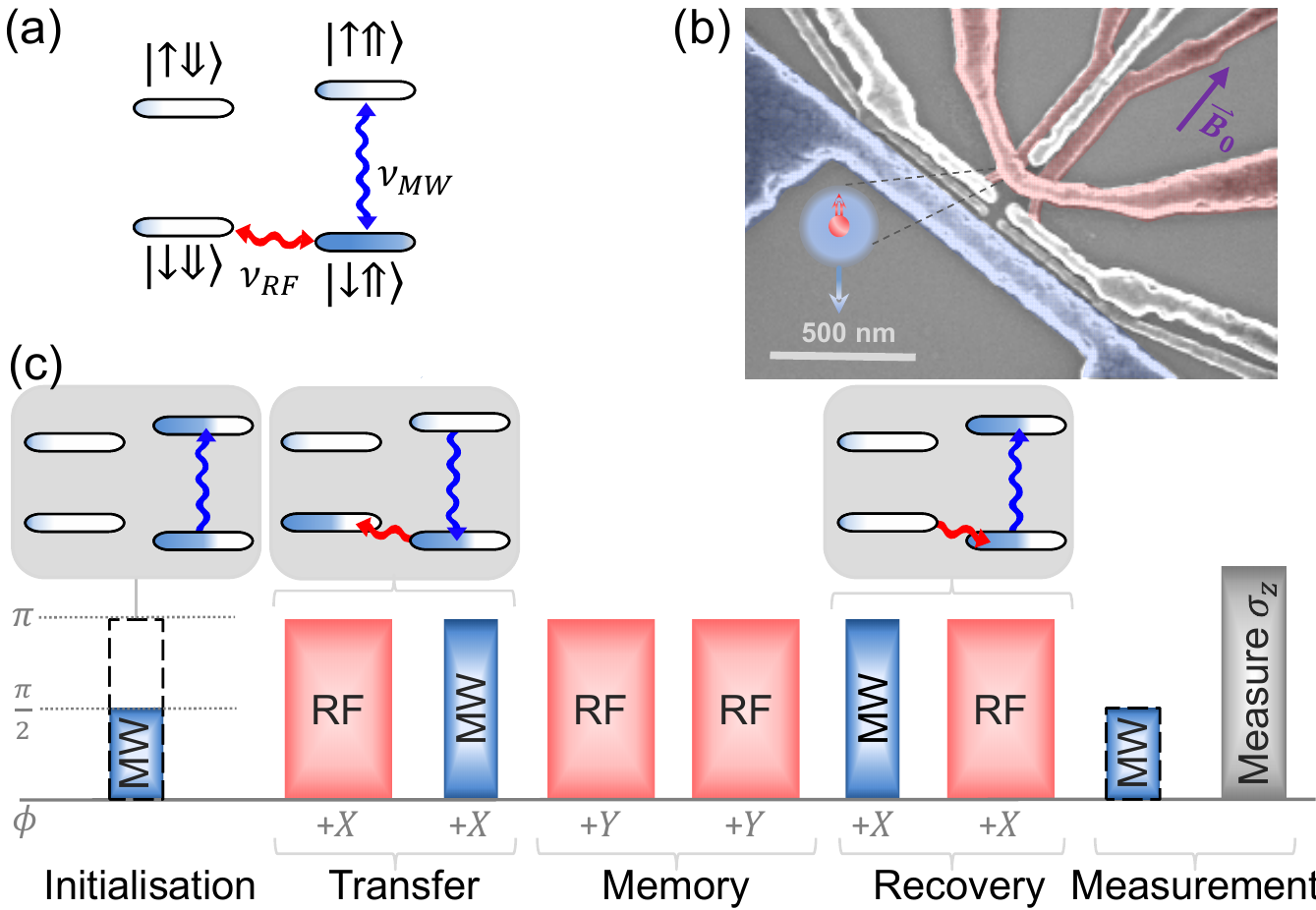}
\caption{\textbf{System schematics and quantum memory sequence}. \textbf{a}, Schematic of the four level, electron-nuclear spin system with the two relevant spin resonance transitions, at frequency $\nu_{RF}$ for the nucleus, and $\nu_{MW}$ for the electron spin. \textbf{b}, Scanning electron micrograph of a device identical to that used in experiments. The single electron transistor (SET) readout device is false-coloured in red, microwae antenna for spin resonance in blue, and tuning gates are uncoloured. \textbf{c}, Memory sequence. A microwave pulse sets the initial quantum state of the electron, ENDOR pulses transfer this coherence to the nucleus, DD pulses extend the coherence during storage. After storage, the ENDOR pulses are reversed to transfer the state back to the electron. An optional tomography MW pulse precedes the electron readout, to allow measurement in an arbitrary basis.}
\label{fig:Figure1}
\end{figure}

\section{Methods and results}
The donor qubit system and supporting nanoelectronics are illustrated in Figure 1(b), which shows a scanning electron micrograph of a device from the same batch as that measured. A single electron transistor (SET), defined electrostatically by aluminium gates on top of a SiO$_2$ insulating layer, is shown in red. Control gates to tune the electrochemical potential of the donor are shown uncoloured, while the broadband microwave antenna \cite{Dehollain2013} used for spin resonance \cite{Pla2012,Pla2013} is shown in blue. The location of the donor atom is indicated by the inset schematic. The device was fabricated on a 900 nm thick epilayer of isotopically enriched $^{28}$Si \cite{Muhonen2014,Itoh2014}, with phosphorus donor atoms implanted \cite{vandonkelaar2015} into the epilayer in a small $90 \times 90$~nm$^2$ window prior to the fabrication of the aluminium nanostructures used for readout and control. The chip was bonded to a high-frequency printed circuit board, mounted in a copper enclosure, and thermalised to the mixing chamber of a dilution refrigerator with electron temperature $T_{el} \sim 100$ mK.  All experiments were performed in a static magnetic field $B_0=1.55$~T, oriented in the plane of the silicon chip along the [110] crystallographic direction.

The Hamiltonian of the system, in frequency units, is ${H=\gamma_e B_0 S_z + \gamma_n B_0 I_z + A \mathbf{S}\cdot\mathbf{I}}$, where $\mathbf{S}$ and $\mathbf{I}$ are the vector Pauli matrices describing the electron and the nuclear spin, respectively, $A=97$ MHz is the electron-nuclear hyperfine interaction for this particular donor \cite{Laucht2015} and $B_0$ is the static magnetic field which defines the $z$-direction. For $B_0=1.55$ T, the eigenstates are almost exactly the tensor product combinations of electron ($\ket{\downarrow}, \ket{\uparrow}$) and nuclear ($\ket{\Downarrow}, \ket{\Uparrow}$) states, as illustrated in the sketch in Figure 1(a). The hyperfine interaction introduces a coupling between the $\ket{\downarrow\Uparrow}$ and $\ket{\uparrow\Downarrow}$ states, which becomes negligible in the limit $\gamma_e B_0 \gg A$ relevant to the present experiment \cite{Kalra2014}. The electron qubit is operated at a microwave frequency $\nu_{MW} = \gamma_e B_0 + A/2 \approx 43$~GHz when the nucleus is in the $\ket{\Uparrow}$ state, while the nuclear spin is operated at radio frequency $\nu_{RF} = A/2 + \gamma_n B_0 \approx 76$~MHz when the electron state is $\ket{\downarrow}$.

\subsection{Qubit measurement and initialisation}

%Initialisation, blip readout etc.
The measurement observable in our system is the expectation value of the $Z$-component of the electron spin $\langle\sigma_z\rangle$. The physical process that gives access to the quantum state of the electron is the energy-dependent tunnelling \cite{Elzerman2004} of the donor-bound electron into the island of the SET \cite{Morello2009,Morello2010}. The large Zeeman splitting between the $\ket{\downarrow}$ and $\ket{\uparrow}$ states ensures that only an electron in the $\ket{\uparrow}$ state can escape the donor, leaving behind a positive charge that shifts the SET bias point and resulting in a current spike detectable in single-shot \cite{Morello2010}. This provides both readout of the electron spin state and initialisation into a $\ket{\downarrow}$ state, since only an electron in the $\ket{\downarrow}$ can tunnel back to the donor. Nuclear initialisation in the $\ket{\Uparrow}$ state is achieved with a sequence of MW and RF $\pi$-pulses \cite{Dehollain2015}. Starting from the $\ket{\downarrow}$ electron state and the nuclear spin in an unknown state, the MW $\pi$-pulse flips the electron to $\ket{\uparrow}$ only if the nucleus is $\ket{\Uparrow}$. In that case, the subsequent RF $\pi$-pulse is off-resonance with the nucleus and leaves it in the target $\ket{\Uparrow}$. Conversely, if the nucleus is initially $\ket{\Downarrow}$, the MW pulse is off-resonance with the electron, and the subsequent RF pulse flips the nucleus from $\ket{\Downarrow}$ to $\ket{\Uparrow}$. With one more electron readout/initialization step, the system is unconditionally prepared in the $\ket{\downarrow\Uparrow}$ electron-nuclear eigenstate before the start of the quantum memory protocol.

\subsection{Pulse sequences and state tomography}
%Note that RF pulses have different powers and lengths. Refer to methods for more
The memory storage and retrieval process is illustrated in Figure 1(c). `Initialisation' describes the preparation of an arbitrary electron state with a microwave pulse of variable length and phase. `Transfer' and `recovery' of this state between electron and nuclear coherences is achieved via the electron-nuclear double resonance (ENDOR) technique \cite{Morton2008}. In the `transfer' stage, the first RF pulse conditionally shifts the $\ket{\downarrow}$ component of the electron spin state onto the $\ket{\Downarrow}$ component of the nucleus, creating a double quantum coherence between $\ket{\uparrow\Uparrow}$ and $\ket{\downarrow\Downarrow}$ if the electron possessed a $\ket{\uparrow}$ component. The subsequent MW $\pi$-pulse conditional on the nuclear $\ket{\Uparrow}$ subspace, translates the $\ket{\uparrow\Uparrow}$ component onto the $\ket{\downarrow\Uparrow}$ state, leaving the nuclear spin with all the quantum information and the electron spin in the $\ket{\downarrow}$ eigenstate. The `transfer' operation of the memory protocol is now complete and the quantum information may be left in the nuclear state for the desired wait time before reversing the order of ENDOR pulses (`recovery' sequence) to bring the quantum information back to the electron state.

For the phase $\phi$ of the RF and MW pulses, we adopt a labelling convention where a $\pi/2$ pulse at $\phi = 0$ produces a $\ket{+X}$ state if the spin started in the $\ket{-Z}$ state. During the storage phase we apply, at a minimum, two nuclear $\pi$-pulses to provide some degree of dynamical decoupling from environmental noise. The phase of these pulses is set at $90^{\circ}$ relative to the phase of the RF transfer pulse. This constitutes a Carr-Purcell-Meiboom-Gill (CPMG) sequence for input states along $X$, or a Carr-Purcell (CP) for input states along $Y$. Following the recovery pulses, the state of the quantum memory is brought back to the electron spin (third inset). A final MW pulse is optionally added to change the electron spin measurement basis.

\begin{figure}
	\centering
	\includegraphics[width=\columnwidth, keepaspectratio = true]{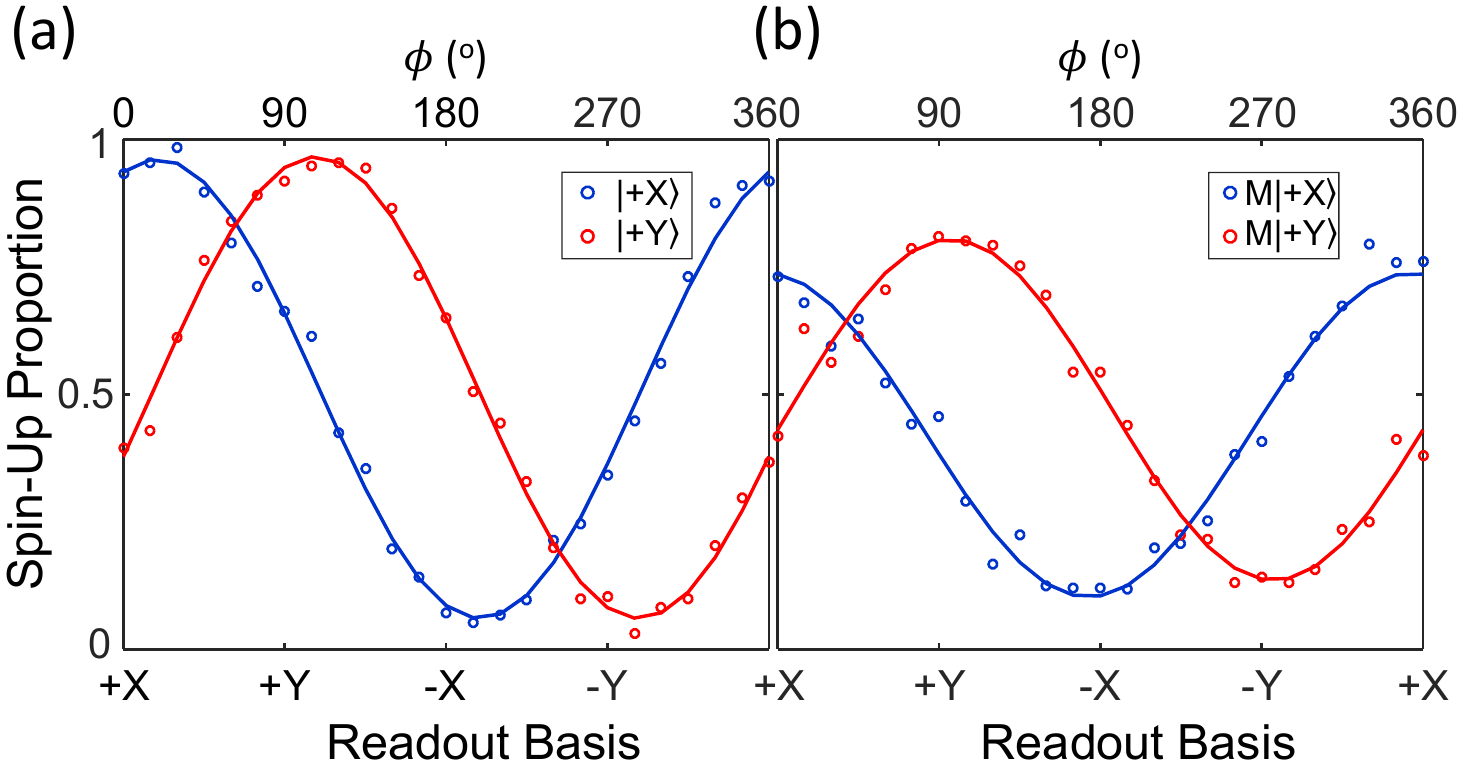}
	\caption{\textbf{State tomography of $\ket{+X}$ and $\ket{+Y}$ in the $XY$ plane}. \textbf{a}, State tomography of an initialised $\ket{+X}$ (blue) and $\ket{+Y}$ (red) state, read out along 24 bases in the $XY$ plane. \textbf{b}, State tomography of a $\ket{+X}$ and $\ket{+Y}$ state following a memory sequence (M) with 196 $\mu$s storage in the nuclear subspace.
		 }
	\label{fig:Figure2}
\end{figure}

% Mention Z-tomography first, show a table of results with error bars
State tomography of a single qubit requires at least three measurement bases \cite{Nielsen2010}. The $Z$ component is the natural measurement basis in our setup, whereas readout of spin components orthogonal to $Z$ is achieved by applying $\frac{\pi}{2}$ rotations about axes in the $XY$ plane (final pulse in Figure 1(c)), mapping the desired readout basis to the $Z$ observable. We used 24 measurement bases along the $XY$ plane, applying a MW $\frac{\pi}{2}$ pulse of varying phase, $\phi\in\{0^{\circ},15^{\circ},30^{\circ},...,345^{\circ}\}$, prior to $Z$-readout.

Figure 2(a) and (b) shows the `$XY$ tomography' before and after 196 $\mu$s of memory storage (1 $\mu$s of free precession and two 97.4 $\mu$s nuclear DD pulses), following the sequence shown in Figure 1(c). Results are shown for an initial state $\ket{+X}=\frac{\ket{\downarrow}+\ket{\uparrow}}{\sqrt2}$ in blue and $\ket{+Y}=\frac{\ket{\downarrow}+i\ket{\uparrow}}{\sqrt2}$ in red. The readout bases $\sigma_{\phi}=\cos(\phi)\sigma_x+\sin(\phi)\sigma_y$ correspond to $+X$, $+Y$, $-X$ and $-Y$ for $\phi$ = 0, 90, 180 and 270$^{\circ}$, respectively. Figure 2(a) shows tomography on the initialised input states, in a sequence consisting only of the `initialisation' and `measurement' stages shown in Figure 1(c). Figure 2(b) shows tomography on the final electron state following the full memory sequence, with 196 $\mu$s storage time.  A sinusoidal fit of the results for each input reveals a maximum signal in a basis approximately aligned with the input superposition, as expected. A small phase shift is noticeable, and is caused by a detuning of order 10~kHz between the electron qubit and the frequency of the MW drive (see Pulse-induced resonance shift). Rather than directly using the $+X$ and $+Y$ results, the amplitude and phase of this sinusoidal fit are used to calculate $\langle\sigma_x\rangle$ and $\langle\sigma_y\rangle$ for tomography. This $XY$ tomography method was designed to minimise errors due to any measurement bias in the tomographic bases chosen.

\begin{figure}
	\centering
	\includegraphics[width=\columnwidth, keepaspectratio = true]{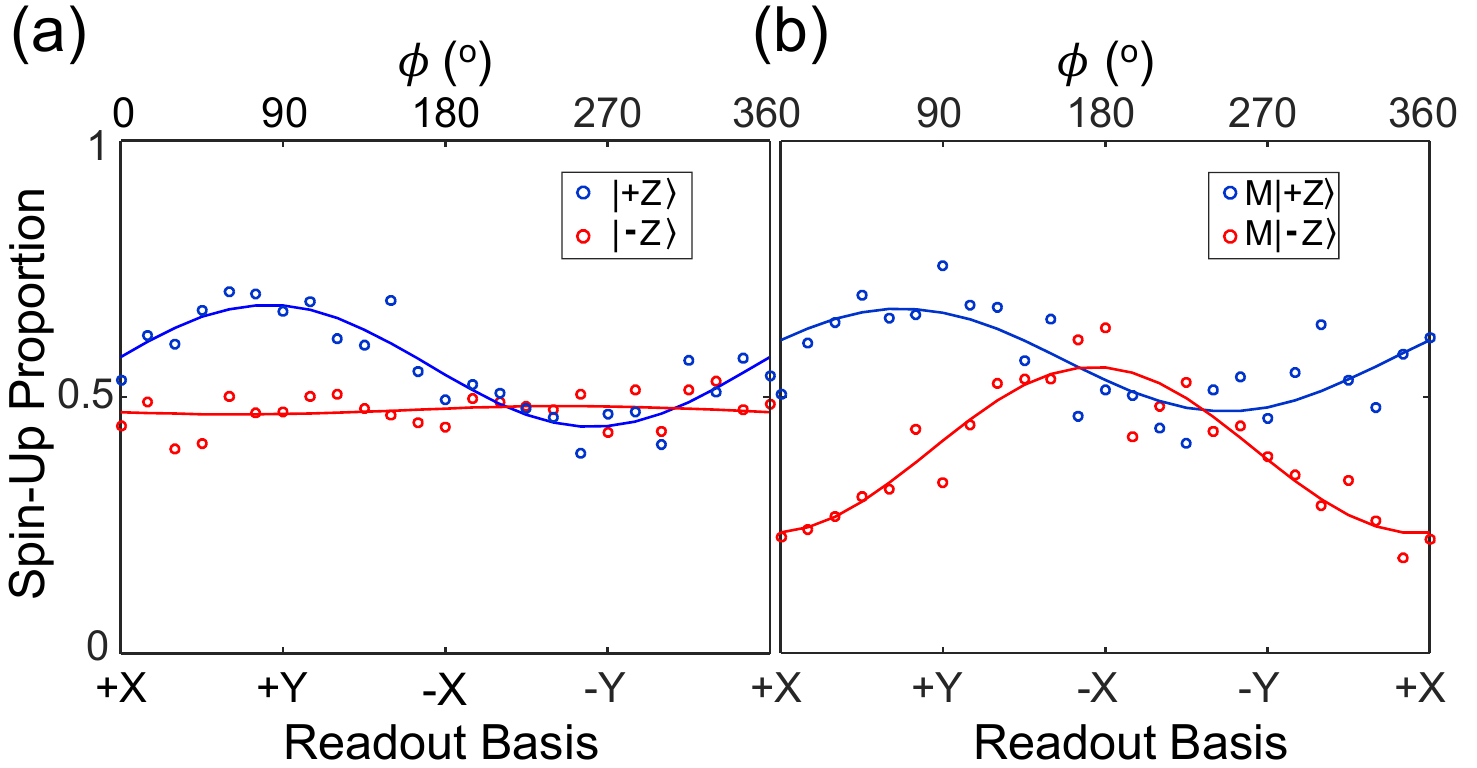}
	\caption{\textbf{State tomography of $\ket{+Z}$ and $\ket{-Z}$ in $XY$ plane}. \textbf{a}, State tomography after initialisation of remaining basis states $\ket{+Z}$ (blue) and $\ket{-Z}$ (red), read out along 24 bases in the $XY$ plane. \textbf{b}, State tomography on the final states $M\ket{+Z}$ and $M\ket{-Z}$ following the same memory sequence (M) as in Figure 2(b).
	}
	\label{fig:Figure3}
\end{figure}

Figure 3 displays the $XY$ plane tomography results for $\ket{+Z}$ and $\ket{-Z}$ states following initialisation (a) and full 196 $\mu$s memory sequence (b). An ideal memory process should produce a horizontal line at 0.5. The oscillations observed in the $XY$-plane for these nominally $Z$-states indicate the presence of detectable $\sigma_x$ or $\sigma_y$ components. These oscillations are due to anomalous time-dependent shifts in the electron spin resonance after the application of an RF pulse, as described below (see Pulse-induced resonance shift). Selecting a single frequency for the sequence forced a compromise between the tuning of the initialisation MW pulse and that of the pulses in the subsequent memory sequence, leaving both slightly detuned from the instantaneous electron resonance frequency. As a result, the initialized states (nominally $\ket{+Z}$ or $\ket{-Z}$) contain spurious $\sigma_x$ and $\sigma_y$ components. However, the initialisation errors can be deconvoluted from the total errors to obtain a `memory fidelity', which represents the fidelity of the storage/retrieval process relative to the actual initialised inputs. %Readout statistics for the final ($Z$) basis are obtained by omitting the change of basis pulse in Figure 1(c) returning to our standard readout observable $\langle\sigma_z\rangle$ and completing the tomographic characterisation of both the input and output state of the 196 $\mu$s memory storage protocol.

% Define purity somewhere
%\subsection{Tomography calculations}
The amplitude of the oscillations in Figure 2 is limited by any non-zero $Z$-component of the state as well as by the maximum visibility of the $Z$ readout. The $X$ and $Y$-components ($\langle\sigma_x\rangle$ and $\langle\sigma_y\rangle$) of the spin were calculated from the amplitude and phase of the curve, with errors determined by the fit. Tomography for the $Z$-component, $\langle\sigma_z\rangle$, is performed separately without change of basis. Using 200 shots per point, the results were averaged over 25 repetitions, with error calculated as the standard deviation of the total distribution.

\begin{figure}
\centering
\includegraphics[width=\columnwidth, keepaspectratio = true]{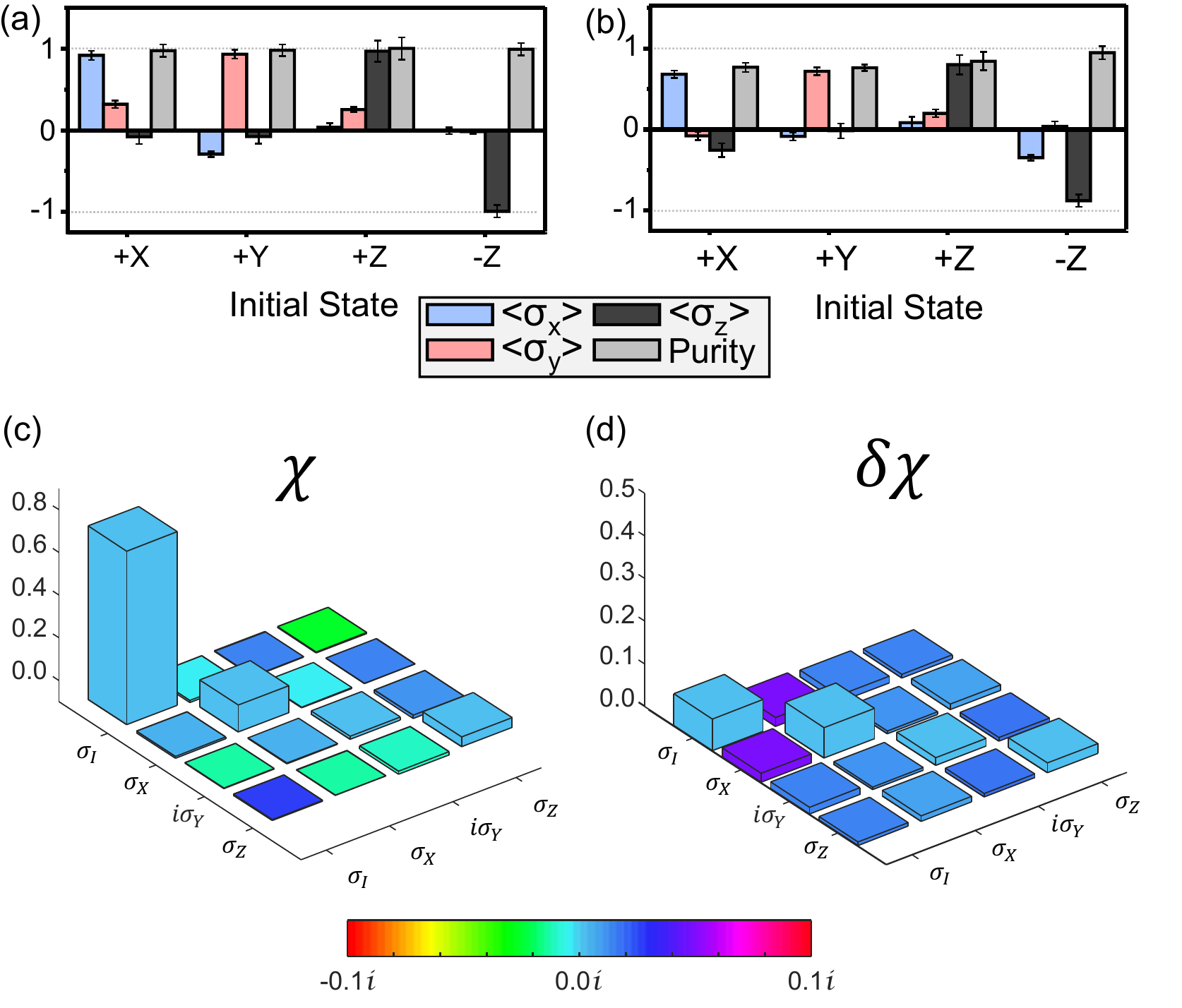}
\caption{\textbf {State and process tomography results for different input states}. \textbf {a}, Before: state tomography results showing expectation values of Pauli spin operators and purity for initialisation of input states $+X$, $+Y$, $+Z$ and $-Z$. \textbf {b}, After: state tomography results for the output states following 196 $\mu$s memory storage and retrieval of the four corresponding input states. \textbf{c}, Process matrix, $\chi$, of the memory protocol with 196 $\mu$s storage. Real components of each process element are indicated by the column heights and imaginary components by the colour. The predominance of the matrix element corresponding to the identity operator indicates successful state preservation, with a process fidelity of 81 $\pm$ 7\%.  \textbf{d}, Estimated standard errors in the process matrix elements calculated via most likelihood estimation procedure and Monte Carlo simulation. The real part of the error is represented by height and the imaginary part by the colour.
}
\label{fig:Figure4}
\end{figure}

Figure 4(a,b) shows the analysis of the state tomography results for the initialised states (a) and output states (b) of the memory process, using four input states ($+X$, $+Y$, $+Z$ and $-Z$) that span the qubit space \cite{Nielsen2010}. The state purity, defined as $Tr(\rho^2)$, is also shown. Comparing the fidelity of the output state with that of the ideal input gives useful information about errors in the storage/retrieval process\cite{Specht2011}, while comparing the input states with the ideal inputs shows errors in state preparation and measurement (SPAM). For a measured state ${\rho}$ and ideal state $\rho_0=\ket{\psi}\bra{\psi}$, the state fidelity is defined as $SF={\langle\psi|\rho|\psi\rangle}$. Table 1 shows the state fidelities for each stored state as well as the fidelity of the initialisation. Averaging over the four stored states gives mean state fidelities $SF_{i} = 98 \pm 9$\% and $SF_{p} = 89 \pm 8$\% for the initialisation and overall process, respectively. Deconvoluting the SPAM errors in the initialisation and read-out from errors in the memory itself implies an average state fidelity $SF_{m} = 91 \pm 12$\% for the memory storage and retrieval itself, comparable with the value obtained in ensemble experiments \cite{Morton2008}.
% Better fidelity formula?

\begin{table}
	
\begin{tabular}{|c|c|c|c|c|c|}
	\hline \textbf{State} & \textbf{+X} & \textbf{+Y} & \textbf{+Z} & \textbf{-Z} & \textbf{Mean} \\
	\hline ${SF_{i}}$ (\%) & 96 ${\pm}$ 3 & 97 ${\pm}$ 3 & 99 ${\pm}$ 6 & 100 ${\pm}$ 4 & 98 ${\pm}$ 9 \\
	\hline ${SF_{p}}$ (\%) & 84 ${\pm}$ 3 & 86 ${\pm}$ 3 & 89 ${\pm}$ 6 & 94 ${\pm}$ 4 & 89 ${\pm}$ 8 \\
	\hline ${SF_{m}}$ (\%) & 87 ${\pm}$ 4 & 89 ${\pm}$ 4 & 91 ${\pm}$ 6 & 94 ${\pm}$ 4 & 91 ${\pm}$ 12 \\
	\hline
\end{tabular}
\label{table:table1}
\caption{State fidelities following initialisation ($SF_i$) and full process ($SF_p$) with $t=196$~$\mu$s storage.
	State fidelities for the memory process ($SF_m$) are calculated as $SF_p/SF_i$}
\end{table}
\subsection{Process tomography}
The storage process was further characterised with quantum process tomography \cite{Nielsen2010}, using the state tomography results on the four basis states. Assuming process linearity, the behaviour of the applied process on these basis states can be characterised as a linear combination of Pauli operators $\sigma_k$ = {$\sigma_I$, $\sigma_X$, $i\sigma_Y$, $\sigma_Z$}. For an input state ${\rho}$, the process is represented as $\xi(\rho) = \sum_{m,n} \sigma_m\rho\sigma_n^\dagger\chi_{mn}$. The matrix $\chi$ is called the process matrix and its diagonal elements $\chi_{kk}$ represent the fidelity of the process relative to the corresponding Pauli operator $\sigma_k$.

The process matrix for a 196 $\mu s$ storage protocol is depicted in Figure 4(c), where the height of each column represents the real part of the matrix element and the imaginary part is described by the colour. Similarly, Figure 4(d) shows the uncertainty in these matrix elements as calculated via maximum likelihood estimation and Monte Carlo analysis \cite{Maurer2012}. The process is dominated by the identity operation, as required for a quantum memory. The matrix element corresponding to row and column $\sigma_I$ indicates a process fidelity $F_p=81 \pm 7$\%. Imaginary and negative components of the matrix are minimal, as expected, and the largest error is a $13\pm7$\% $\sigma_x$-like component. Process tomography on the initialised inputs reveals an initialisation fidelity $F_i=88\pm6$\%, with a 7\%  $\sigma_x$-like error. Adjusting the process fidelity for this initialisation error gives an estimated fidelity for the memory process itself $F_m=F_p/F_i=92\pm11$\%, comparable to the value obtained with defects in diamond \cite{Fuchs2011,Maurer2012}. The remaining errors are largely due to a systematic shift in the instantaneous ESR frequency induced by application of RF pulses, causing errors in rotation angle and phase during both the initialisation and the memory sequences (see Pulse-induced resonance shift below).
% Andrea: why 85% when the SF's for init were so high? - It's a bit weird, but the non-diagonal terms in the X_i are just quite numerous.

%\subsection{Error calculations}
Estimating the errors in the process matrix elements requires repeated applications of the maximum likelihood procedure to multiple sets of tomographic data with real or simulated errors, as described e.g. in Refs. \cite{Maurer2012,Obrien2004}. Gaussian noise was added to the density matrix elements based on the measurement uncertainties, simulating output states within the error bars of the measurement. For each simulation, a process matrix is generated that has the maximum likelihood of producing these outputs via least squares minimisation. An extra constraint is added to maximise the hermiticity and positivity of each $\chi$ \cite{Maurer2012}. A Monte Carlo distribution of 20,000 process matrices and sets of state fidelities were generated. The standard deviations in the set of 20,000 results were used to estimate the error in the calculated fidelities.

\subsection{Pulse-induced resonance shift}
The power of the RF pulses applied during the transfer and recovery stages of the memory sequence needs to be high enough to ensure that the transfer time is short compared to the electron dephasing time ($T_{2e}^{\ast}\approx 160$~$\mu$s) \cite{Muhonen2014}. In the present experiment, we used typical RF powers of $-1$~dBm at the source, which corresponds to approximately $-15$~dBm at the sample. Following the application of these high-power RF pulses it was found that the electron spin resonance frequency $\nu_{e}$ exhibits a transient shift, as shown in Figure 5. The peak frequency detuning is typically $|\nu_e-\nu_{\rm MW}| \sim 10$~kHz and lasts on the order of 200 $\mu$s. Plotting on the same graph the time dependence of $|\nu_e-\nu_{\rm MW}|$ (blue) and of the SET current (red) following an RF pulse shows that the two are correlated. However, we found no causal link between the SET current and the frequency detuning, i.e. deliberately running a current through the SET does not shift $\nu_{e}$. In this test, the RF pulse was tuned completely off-resonance and any frequency dependence of the effect was negligible over the MHz range of the RF source used. The shift was found to be positive for both ESR frequencies (data not shown), implying an effective shift in the Zeeman term of the Hamiltonian, rather than in the hyperfine coupling. The physical origin of this resonance shift is currently unknown. 

In the present experiment, this pulse-induced resonance shift results in a different instantaneous electron resonance frequency during the `initialization' MW pulse (which precedes any RF pulse) and during the ENDOR sequences, preventing us from achieving perfect initialization and/or state transfer. In the practice, we chose a compromise where the MW source is slightly detuned from the instantaneous electron resonance under both conditions. We anticipate that improved memory fidelities will be achieved once the pulse-induced resonance shift is understood and mitigated.

\begin{figure}
	\centering
	\includegraphics[width=\columnwidth, keepaspectratio = true]{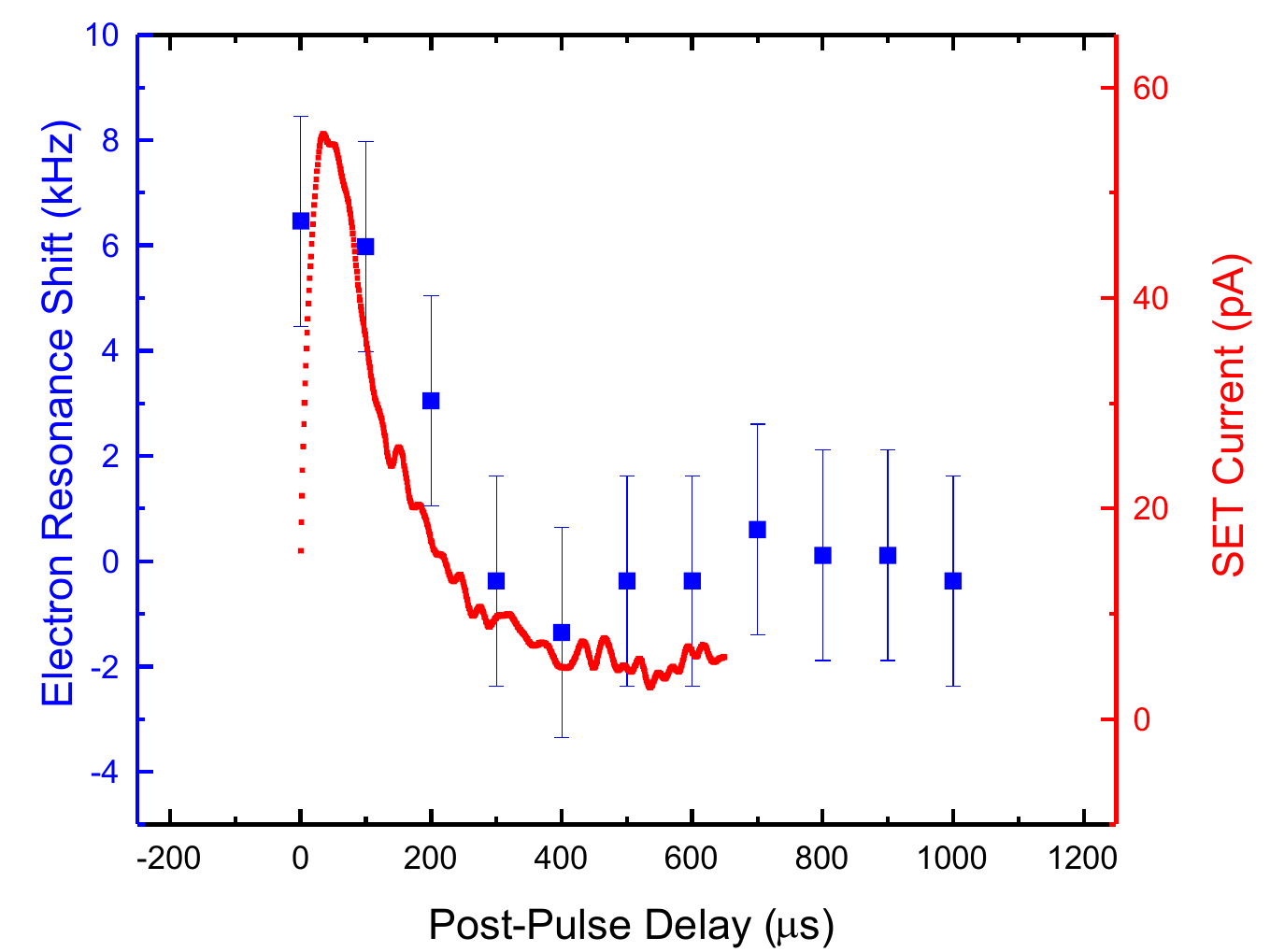}
	\caption{\textbf{Pulse-induced resonance shift}. Detuning of the ESR frequency (blue, left axis) as a function of the delay time following an off-resonant RF pulse. The instantaneous electron resonance frequency was extracted from a Ramsey interference experiment, using MW $\pi/2$ pulses shifted by 15 kHz from the steady-state resonance. The instantaneous SET current (red, right axis) following the same RF pulse shows a time dependence that correlates with the ESR frequency shift.
	}
	\label{fig:Figure5}
\end{figure}

% Used RF pulses of different power due to this
\subsection{Memory lifetime}
As a result of the long coherence time of the nuclear spin, the process fidelity decays slowly with storage time. For instance we measure $F_p = 74 \pm 3$\% after 10 ms, corresponding to a memory fidelity $F_m\approx87$\%. High-fidelity storage can be extended to longer times by dynamically decoupling the nuclear spin. Figure 6(a) shows the probability of recovering an initial $X$-state as a function of storage time, for different numbers of DD pulses applied to the nucleus during storage. This can be thought of as the CPMG coherence time, $T^{\rm CPMG}_2$ of the memory storage process. Figure 6(b) compares this memory $T^{\rm CPMG}_2$ (black) to nuclear CPMG coherence times (red) with varying number $N$ of applied CPMG pulses. Both values approach 80 ms for $N=256$ pulses, with the memory following a similar trend to that of the neutral nucleus. Electron CPMG coherence times, shown in blue, are shorter than the nuclear and memory times for small $N$, but tend to approach them for higher number of CPMG pulses. The values of  $T^{\rm CPMG}_2$ as a function of $N$ show increase in proportion to $N^{0.75}$, $N^{0.28}$ and $N^{0.36}$ for the electron, nucleus and memory, respectively.

\begin{figure}
	\centering
	\includegraphics[width=\columnwidth, keepaspectratio = true]{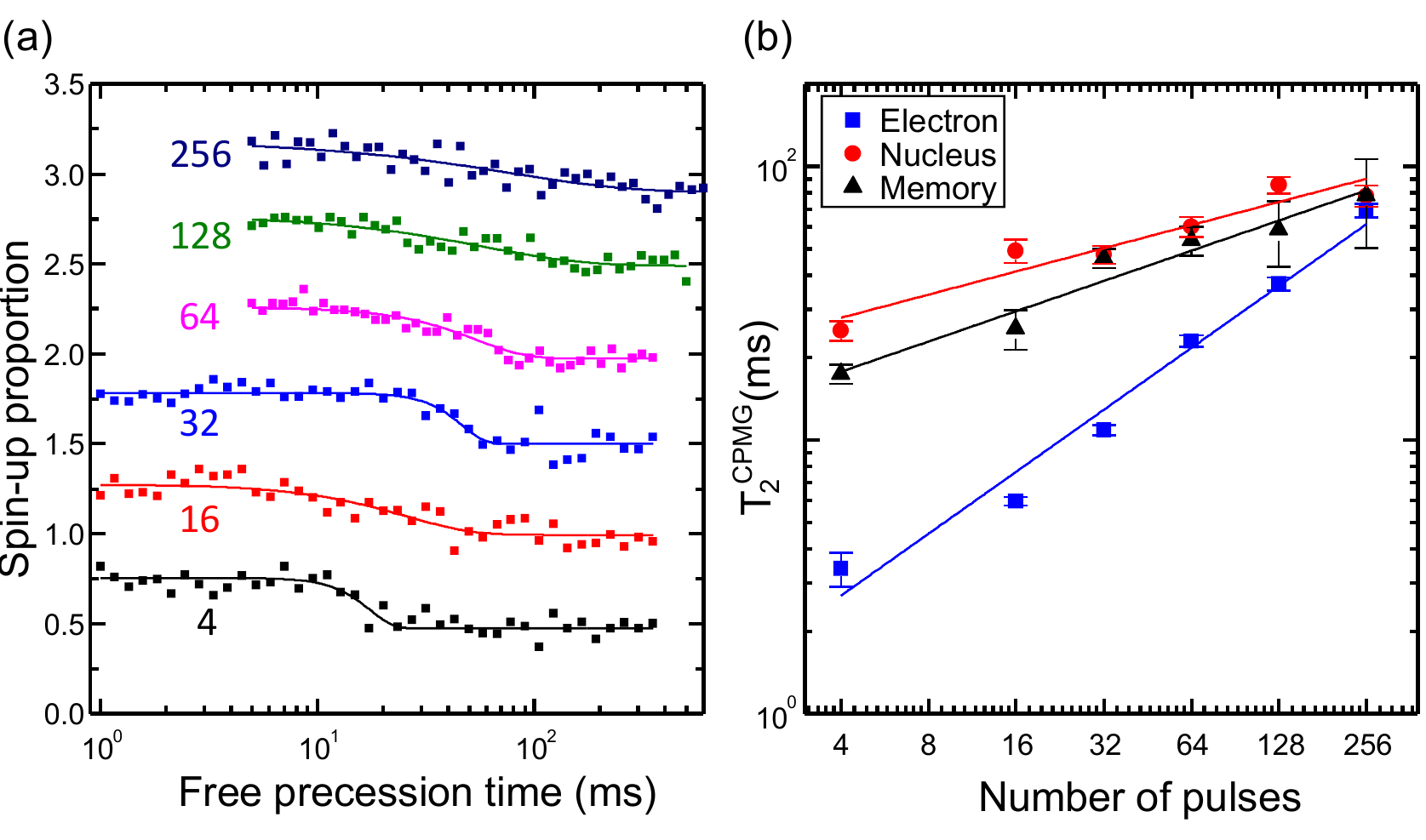}
	\caption{\textbf{Memory storage times}. \textbf{a}, Coherence decay for a $+X$ input state, read out in the $+X$ basis after varying storage time, for different number $N$ of CPMG dynamical decoupling pulses. The data is fitted by a stretched exponential function of the form $y=y_0+K\exp(-(\tau/T_2)^{\alpha})$. The  curves are offset in increments of 0.5 for clarity. \textbf{b}, Comparison of coherence times (from fits) for the memory protocol, the neutral nucleus and the electron, as a function of the number of CPMG pulses.
	}
	\label{fig:Figure6}
\end{figure}

The different trend of $T^{\rm CPMG}_2$ between electron, nucleus and memory as a function of $N$ is not fully understood, but is suspected to be related to the pulse-induced resonance shift described above. We note indeed that the neutral nucleus coherence times reported here are much longer than those measured in an earlier experiment on the same device \cite{Muhonen2014}, where even higher RF powers were used. This suggests that there might be a pulse-induced resonance shift on the nuclear spin as well -- too small to be detected directly in a Ramsey experiment as we managed to do for the electron (Figure 5), but large enough to perturb the coherent evolution of the nuclear spin. While this can be mitigated to some extent by reducing the RF power, the memory protocol requires nuclear $\pi$-pulses shorter than the electron dephasing time  $T_{2e}^{\ast}\approx 160$~$\mu$s \cite{Muhonen2014} for faithful transfer of the electron state to the nucleus. We found the best results by choosing memory transfer RF $\pi$-pulses of duration 50 $\mu$s, and lower-power $\pi$-pulses (97 $\mu$s) applied for DD during the storage phase.

\section{Conclusion}
In conclusion, we have experimentally demonstrated a nuclear spin quantum memory using a single P atom in silicon. The memory coherence time (up to 80 ms) and the memory process fidelity (better than 90\%) are comparable to those found in nitrogen-vacancy centres in diamond \cite{Fuchs2011,Maurer2012}, trapped ions \cite{Specht2011} and P-atom ensembles \cite{Morton2008}. We have identified a likely cause of the coherence and fidelity limitations, namely a shift in the instantaneous electron resonance frequency after applying RF pulses. Future work will focus on understanding and eliminating this pulse-induced frequency shift, in order to demonstrate quantum memory fidelities approaching the values of gate fidelity ($\gg 99$\%) \cite{Muhonen2015,Dehollain2016} already observed for the operation of the electron and nuclear spin qubits individually.

\section*{Acknowledgements}
This research was funded by the Australian Research Council Centre of Excellence for Quantum Computation and Communication Technology (project number CE11E0001027), the US Army Research Office (W911NF-13-1-0024) and the Commonwealth Bank of Australia. We acknowledge support from the Australian National Fabrication Facility, and from the laboratory of Prof Robert Elliman at the Australian National University for the ion implantation facilities. The work at Keio has been supported in part by KAKENHI (S) No. 26220602, Core-to-Core Program by JSPS, and Spintronics Research Network of Japan.
%\newline \\
%\textbf{Authors Contributions} S.F., S.S. and A.M. designed the experiments. S.F. and S.S. performed the measurements and analysed the results. D.N.J. and J.C.M. designed the P implantation experiments. F.E.H. fabricated the device with A.S.D.'s supervision and R.K.'s assistance. K.M.I. prepared and supplied the $^{28}$Si epilayer wafer. S.F., S.S. and A.M. wrote the manuscript, with input from all coauthors.
%\newline \\
%The authors declare no competing financial interests.
%\newline \\
%Correspondence should be addressed to S.F. (slf@unsw.edu.au) \\
%or A.M. (a.morello@unsw.edu.au). \pagebreak

\section*{References}
%\bibliographystyle{iopart-num}
%\bibliography{Freer_QST-refs}

\providecommand{\newblock}{}

\end{document}